\documentclass[twolcolumn,apj]{emulateapj}

\usepackage{amsmath}
\pdfoutput=1

\begin{document}

\title{PAPER-64 Constraints On Reionization II: The Temperature Of The $z=8.4$ Intergalactic Medium}

\author{Jonathan C. Pober\altaffilmark{1,19}, 
Zaki S. Ali\altaffilmark{2}, 
Aaron R. Parsons\altaffilmark{2,3},
Matthew McQuinn\altaffilmark{4},
James E. Aguirre\altaffilmark{5},
Gianni Bernardi\altaffilmark{6,7,8},
Richard F. Bradley\altaffilmark{9,10,11}, 
Chris L. Carilli\altaffilmark{12,13},
Carina Cheng\altaffilmark{2},
David R. DeBoer\altaffilmark{3},
Matthew R. Dexter\altaffilmark{3},
Steven R. Furlanetto\altaffilmark{14},
Jasper Grobbelaar\altaffilmark{6},
Jasper Horrell\altaffilmark{6},
Daniel C. Jacobs\altaffilmark{15,19},
Patricia J. Klima\altaffilmark{10},
Saul A. Kohn\altaffilmark{5},
Adrian Liu\altaffilmark{2,16},
David H. E. MacMahon\altaffilmark{3},
Matthys Maree\altaffilmark{6},
Andrei Mesinger\altaffilmark{17},
David F. Moore\altaffilmark{5},
Nima Razavi-Ghods\altaffilmark{13}, 
Irina I. Stefan\altaffilmark{13},
William P. Walbrugh\altaffilmark{6},
Andre Walker\altaffilmark{6},
Haoxuan Zheng\altaffilmark{18}
}
\altaffiltext{1}{Physics Dept., U. Washington, Seattle, WA}
\altaffiltext{2}{Astronomy Dept., U. California, Berkeley, CA}
\altaffiltext{3}{Radio Astronomy Lab., U. California, Berkeley, CA}
\altaffiltext{4}{Astronomy Dept., U. Washington, Seattle, WA}
\altaffiltext{5}{Dept. of Physics and Astronomy, U. Pennsylvania, Philadelphia, PA}
\altaffiltext{6}{Square Kilometre Array South Africa (SKA SA), Pinelands, South Africa}
\altaffiltext{7}{Dept. of Physics and Electronics, Rhodes U., Grahamstown, South Africa}
\altaffiltext{8}{Harvard-Smithsonian Center for Astrophysics, Cambridge, MA}
\altaffiltext{9}{Dept. of Electrical and Computer Engineering, U. Virginia, Charlottesville, VA}
\altaffiltext{10}{National Radio Astronomy Obs., Charlottesville, VA}
\altaffiltext{11}{Dept. of Astronomy, U. Virginia, Charlottesville, VA}
\altaffiltext{12}{National Radio Astronomy Obs., Socorro, NM}
\altaffiltext{13}{Cavendish Lab., Cambridge, UK}
\altaffiltext{14}{Dept. of Physics and Astronomy, U. California, Los Angeles, CA}
\altaffiltext{15}{School of Earth and Space Exploration, Arizona State U., Tempe, AZ}
\altaffiltext{16}{Berkeley Center for Cosmological Physics, Berkeley, CA}
\altaffiltext{17}{Scuola Normale Superiore, Pisa, Italy}
\altaffiltext{18}{Dept. of Physics, Massachusetts Institute of Technology, Cambridge, MA}
\altaffiltext{19}{National Science Foundation Astronomy and Astrophysics Postdoctoral Fellow}

\begin{abstract}
We present constraints on both the kinetic temperature of the intergalactic medium (IGM) at $z=8.4$,
and on models for heating the IGM at high-redshift with X-ray emission from the first collapsed 
objects.  These constraints are derived
using a semi-analytic method to explore the new measurements of the 21~cm power
spectrum from the Donald C. Backer Precision Array for Probing the Epoch of Reionization
(PAPER),
which were presented in a companion paper, \cite{ali_et_al_2015}.
Twenty-one~cm power spectra with amplitudes of hundreds of mK$^2$
can be generically produced if the kinetic temperature of the 
IGM is significantly below the temperature
of the Cosmic Microwave Background (CMB); as such, the new results from
PAPER place lower limits on the IGM temperature at $z=8.4$.  
Allowing for the unknown ionization state of the IGM, our measurements
find the IGM temperature to be above $\approx~5$~K for neutral fractions between 10\% and 85\%, 
above $\approx~7$~K for neutral fractions between 15\% and 80\%, 
or above $\approx~10$~K for neutral fractions between 30\% and 70\%.
We also calculate the heating of the IGM that would be provided by the observed high redshift
galaxy population, and find that for most models, these galaxies are sufficient to bring the IGM
temperature above our lower limits.  However, there are significant ranges of parameter
space that could produce a signal ruled out by the PAPER measurements;
models with
a steep drop-off in the star formation rate density at high redshifts or 
with relatively low values for the X-ray to star formation rate
efficiency of high redshift galaxies
are generally disfavored.
The PAPER measurements are consistent with (but do not constrain) 
a hydrogen spin temperature above the CMB temperature, a situation which we find to be
generally predicted if galaxies fainter than the 
current detection limits of optical/NIR surveys are included in calculations of X-ray heating.

\end{abstract}

\keywords{reionization, dark ages, first stars --- intergalactic medium --- galaxies: high-redshift}

\section{Introduction}

Up until very recently, observational cosmology has lacked a direct probe
of the conditions of the Universe between the epoch
of recombination and the birth of modern galaxies.
Recent observations with the Hubble Space Telescope 
have found nearly a thousand putative galaxies at a redshift of 7 or above
(see \citealt{bouwens_et_al_2014} for a recent compilation), using the
Lyman break drop-out technique. Narrow band searches for Lyman-$\alpha$
emitters have also proven successful at finding high-redshift galaxies
\citep{tilvi_et_al_2010,ouchi_et_al_2010,hibon_et_al_2011}.
However, these observations still principally detect the rare, most luminous
galaxies, and cannot probe the population of more numerous, fainter
galaxies.  

One of the biggest unsolved questions in cosmology and galaxy formation
is the issue of how early galaxies feed back into the Universe
and influence the formation of the next generation of galaxies.
From a combination of Cosmic Microwave Background (CMB; \citealt{larson_et_al_2011,planck_2015_13})
observations
and quasar absorption spectra measurements at high redshift
\citep{fan_et_al_2006,mcgreer_et_al_2015}, we know that ultraviolet photons from the first
luminous objects reionized the intergalactic medium (IGM) somewhere between
redshifts $\approx 6$ and 15.
However, it has been determined that the observed high-redshift galaxy population
cannot produce enough ionizing photons to 
complete the reionization of the Universe
before $z=6$
\citep{choudhury_et_al_2008,kuhlen_and_faucher_giguere_2012,finkelstein_et_al_2012,robertson_et_al_2013}.
Therefore, to understand cosmic reionization, heating, and other
feedback effects on the IGM, we require a probe of 
global conditions which captures the impact of the unobservable low-mass
galaxies.

The 21~cm line of neutral hydrogen offers such a probe.  By observing the 21~cm
signal as a function of redshift, one can potentially trace the evolution of
ionization, temperature, and density fluctuations on a cosmic scale
(for reviews of 21~cm cosmology, see \citealt{furlanetto_et_al_2006},
\citealt{morales_and_wyithe_2011} and \citealt{pritchard_and_loeb_2012}).
At present, telescopes such as
the LOw Frequency ARray (LOFAR; \citealt{yatawatta_et_al_2013,van_haarlem_et_al_2013})\footnote{http://www.lofar.org},
the Murchison Widefield Array (MWA; \citealt{lonsdale_et_al_2009,tingay_et_al_2013,bowman_et_al_2013})\footnote{http://www.mwatelescope.org},
and the Donald C. Backer Precision Array for Probing the Epoch of Reionization
(PAPER; \citealt{parsons_et_al_2010})\footnote{http://eor.berkeley.edu}
are conducting lengthy observational campaigns
to detect the spatial power spectrum of 21~cm fluctuations from the
Epoch of Reionization (EoR).  

Initial measurements from a 32-element PAPER instrument in 2011 were recently
used to place an upper limit on the 21~cm power spectrum at redshift 7.7
at a wavenumber of $k = 0.27~h\rm{Mpc}^{-1}$
\citep{parsons_et_al_2014}.  This upper limit was
stringent enough to` place constraints on the temperature of the IGM,
requiring some mechanism for heating the intergalactic gas, and
ruling out a universe 
in which the thermal evolution of the IGM was purely adiabatic
since decoupling from the CMB.
The goal of the present work is to expand on this analysis by using the 
more stringent upper limit from \cite{ali_et_al_2015} (hereafter, ``Paper I") and
by using a semi-analytic method to model the signal,
allowing for a more complete exploration of the parameter space.
This approach improves over
\cite{parsons_et_al_2014} and Paper I, which constrained the spin temperature of the IGM
by treating the brightness contrast between the spin and CMB temperatures
as a multiplicative scalar on the amplitude of an analytic ``patchy" reionization power spectrum.
We review the measurements of Paper I in \S\ref{sec:data},
and outline our methodology in \S\ref{sec:methods}.
We present our constraints on the IGM temperature in \S\ref{sec:results},
and discuss their physical implications in \S\ref{sec:discussion}.
We conclude in \S\ref{sec:conclusion}.  Unless otherwise stated,
all calculations assume a flat $\Lambda$CDM universe with $\Omega_m = 0.27,\
\Omega_\Lambda = 0.73$, $n_s = 0.96$, $\sigma_8 = 0.82$, and $h = 0.7$.

\section{Data and Measurements}
\label{sec:data}

The power spectrum measurements presented in Paper I represent a substantial improvement
over the previous limits from PAPER in \cite{parsons_et_al_2014}.
While the previous measurements placed an upper limit of $\Delta^2 (k) \leq (41\,\textrm{mK})^{2}$ at 
$k=0.27~h{\rm Mpc}^{-1}$ and $z=7.7$, the new limits from Paper I are significantly lower:
$\Delta^2(k) \leq (22.4~{\rm mK})^2$ over the range $0.15<k<0.5~h\rm{Mpc}^{-1}$ at $z=8.4$.
This factor of 2 reduction (a factor of 4 in the temperature-squared units of the power
spectrum) is the result of a large number of improvements over the previous analysis.
The measurements in Paper I come from a 64-element PAPER array, as opposed to the 32-element
array used in \cite{parsons_et_al_2014}.  There are also several significant changes to the
data processing and analysis compared with that in \cite{parsons_et_al_2014}.  
First, the application
of the Omnical\footnote{https://github.com/jeffzhen/omnical}\ redundant calibration
package \citep{zheng_et_al_2014} to the visibility data substantially improves
the calibration and reduces the variance among measurements from physically
redundant baselines.  Second, the application of optimal fringe rate filtering has the effect
of upweighting higher signal-to-noise regions on the sky and can limit the contamination
from foreground emission at the northern and southern horizons.  Finally, an improved
power spectrum estimator using the optimal quadratic estimator formalism 
\citep{tegmark_1997,liu_and_tegmark_2011,dillon_et_al_2013a} in conjunction with the delay
spectrum approach \citep{parsons_et_al_2012b} is used to downweight spectral eigenmodes that
show significant foreground contamination.  In a significant change from the
\cite{parsons_et_al_2014} analysis, only the covariance between frequency channels within
a single baseline's measurements is used in the weighting, as opposed to the
covariance removal techniques that focused on inter-baseline covariance.  See Paper I for
a detailed description of each of these three new techniques.  

\begin{figure*}[htbp!]
\centering
\includegraphics[width=6.5in]{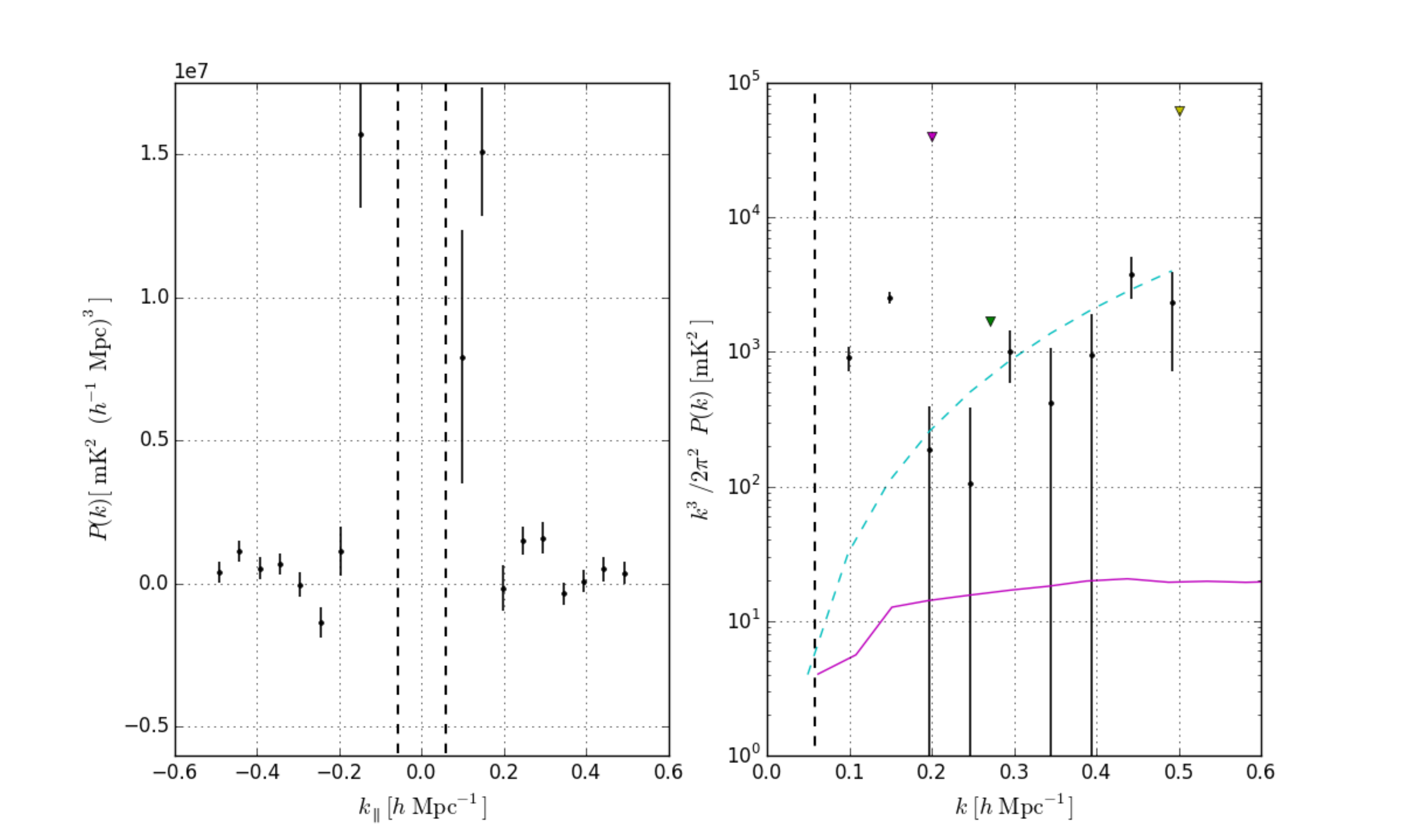}
\caption{
\emph{Left:} The measured $P(k)$ in units of $\mathrm{mK}^2(h^{-1}\mathrm{Mpc})^3$ over both
positive and negative line-of-sight wavenumber $k_{\parallel}$. \emph{Right:} 
The dimensionless power spectrum $\Delta^2(k) = \frac{k^3}{2\pi^2} P(k)$ in units of mK$^2$ 
versus $|k|$.  In both panels, black
points represent the new measurements with 2$\sigma$ error bars derived from bootstrapping,
while black dashed lines represent the nominal horizon limit to flat
spectrum foreground emission.  Also shown in the right hand panel
are the expected theoretical $2\sigma$ upper bounds of a noise-dominated
measurement (dashed cyan), a model 21~cm power spectrum at
50\% ionization from \cite{lidz_et_al_2008} (magenta), and three previous upper limits
on the 21~cm signal: the Giant Metrewave Radio Telescope measurement at $z=8.6$
(\citealt{paciga_et_al_2013}; yellow triangle); the MWA measurement at $z=9.5$ 
(\citealt{dillon_et_al_2013b}; purple triangle); and the \cite{parsons_et_al_2014}
measurement (green triangle).
}
\label{fig:data_pspec}
\end{figure*}

The new power spectrum measurements from Paper I are shown in Figure \ref{fig:data_pspec}
(equivalent to Figure 18 in Paper I).
The left hand panel shows $P(k)$ in $\mathrm{mK}^2(h^{-1}\mathrm{Mpc})^3$, while the right hand panel
plots the dimensionless power spectrum $\Delta^2(k)$ in mK$^2$.  In both panels, black
points represent the new measurements with 2$\sigma$ error bars derived from bootstrapping,
while vertical dashed black lines 
represent the nominal horizon limit, beyond which 
there should be no contamination from flat spectrum foreground emission
\citep{parsons_et_al_2012b,pober_et_al_2013}.  Also shown in the right hand panel
are the expected theoretical 2$\sigma$ noise limit (i.e. 95\% of points
should fall under this line if the measurements are consistent with thermal noise;
dashed-cyan), a model 21~cm power spectrum at
50\% ionization from \cite{lidz_et_al_2008} (magenta), and three previous upper limits
on the 21~cm signal: the Giant Metrewave Radio Telescope measurement at $z=8.6$
(\citealt{paciga_et_al_2013}; yellow triangle); the MWA measurement at $z=9.5$ 
(\citealt{dillon_et_al_2013b}; purple triangle); and the \cite{parsons_et_al_2014}
measurement (green triangle).

\section{Methodology}
\label{sec:methods}

As a way of placing these results in the context of a large and uncertain 
parameter space, we identify two parameters as
being the dominant contributors to 21~cm power spectra with amplitudes at the level of
the PAPER constraints:
the average spin temperature of the emitting (i.e. neutral) gas 
--- which at these redshifts is set by the kinetic temperature of the gas ---
and the average neutral fraction.
To understand the physical conditions under which these two parameters
drive the 21~cm power spectrum, it is worthwhile to keep the brightness
temperature contrast between the 21~cm signal and the CMB, $\delta T_b$,
in mind:
\begin{align}
\label{eq:deltaTb}
\delta T_b(\nu) \approx &\ 9 x_{\rm HI} (1+\delta) (1+z)^\frac12 \nonumber \\
\times&\left[1 - \frac{T_{\rm CMB}(z)}{T_S}\right] \left[\frac{H(z)/(1+z)}{\mathrm{d}v_{\parallel}/\mathrm{d}r_{\parallel}}\right] \mathrm{mK},
\end{align}
where $x_{\rm HI}$ is the global neutral hydrogen fraction, $z$ is the
redshift, $T_{\rm CMB}$ is the temperature of the cosmic microwave background,
$T_S$ is the spin temperature, $H(z)$ is the Hubble parameter, and
$\mathrm{d}v_{\parallel}/\mathrm{d}r_{\parallel}$ is the gradient
of the proper velocity along the line of sight \citep{furlanetto_et_al_2006}.
It is worth explicitly stating that all the terms in Equation \ref{eq:deltaTb}
can have different values at different spatial locations in the universe.
When we refer to the \emph{morphology} of, e.g., the ionization or spin temperature field,
we are referring to the spatial distribution of the fluctuations in these quantities.
We often quantify the statistics of these fluctuations in cosmological Fourier space using a power
spectrum.
If we define a fractional brightness temperature perturbation,
$\delta_{21}(\vec{x}) \equiv[\delta T_b(\vec{x}) - \bar{\delta T_b}]/\bar{\delta T_b}$,
the power spectrum, $P(\vec{k})$, is given by the ensemble average of the square of 
the spatial Fourier transform of this perturbation:
\begin{equation}
\left<\tilde\delta_{21}(\vec{k})\tilde\delta_{21}(\vec{k}')\right> \equiv (2\pi)^3 \delta(\vec{k}-\vec{k}') P(\vec{k}),
\end{equation}
where the unsubscripted $\delta$ is a Dirac delta function.

Using these relations as a framework, we can now discuss the impact of our
two principal parameters on the 21~cm power spectrum.

\subsection{Spin Temperature}

In the brightness temperature $\delta T_b$, the spin temperature enters
as a ratio with the CMB temperature: $[1 - T_{\rm CMB}(z)/T_S]$.  If the spin
temperature is much larger than the CMB temperature, this term saturates at a value of 1.  
It is often assumed during reionization that the spin
temperature is already very large (e.g. \citealt{furlanetto_2006}, \citealt{pritchard_and_loeb_2008}, \citealt{pober_et_al_2014}).
It is thought that
the emission of ultraviolet photons from the first luminous objects couples
the spin temperature to the kinetic gas temperature
field through the Wouthuysen-Field effect \citep{wouthuysen_1952,field_1958}.\footnote{
Using Equation 7 from \cite{mcquinn_and_oleary_2012}, we estimate that a star formation
rate density of $2.5~\times~10^{-3}~{\mathrm{M}_\odot \mathrm{Mpc}^{-3} \mathrm{yr}^{-1}}$
is necessary for the Wouthuysen-Field effect to couple the spin and color temperatures
in the IGM by $z=8.4$.  The observed high redshift star formation rate density is nearly
an order of magnitude higher than this value \citep{bouwens_et_al_2014,mcleod_et_al_2014},
making the assumption that the spin temperature is equivalent to the kinetic temperature
of the gas a valid one.}
And, in most models of early star and
galaxy formation, the kinetic gas temperature has been raised to a very high
level through heating from X-rays from the first high-mass X-ray binaries (HMXBs).  
(Recent work by \citealt{fialkov_et_al_2014} has called this last statement into question,
motivating the ``cold" reionization scenarios we consider here.)
However, it is clear that a very low value of the spin temperature --- as can
occur if X-ray heating is inefficient --- 
can make the relevant term in the 21~cm brightness temperature
(Equation \ref{eq:deltaTb}) large
and negative, meaning the hydrogen gas is seen in absorption relative to
the CMB.
It is worth stressing that in our model $T_S$ corresponds to the
mass-averaged spin temperature.
Like the ionization and density fields, the spin temperature also fluctuates spatially
and contributes to the overall 21~cm power spectrum.  As we discuss below, we do not vary
the morphology of the spin temperature fluctuations in our simulations, but rather vary only
the total intensity of heating, effectively scaling the global temperature field.

\subsection{Neutral Fraction}
The evolution of the shape of 21~cm power spectrum as reionization proceeds is
largely driven by the evolution in the neutral fraction, $x_{\rm HI}$,
and its spatial fluctuations, $\delta {x_{\rm HI}}$.
Simulations have shown
that neutral fraction largely serves as a time coordinate during reionization;
put another way, the shape (and, to a lesser degree, amplitude) of the 
power spectrum can largely be mapped one-to-one to the global neutral fraction,
independent of the redshift at which the neutral fraction actually occurs
\citep{mcquinn_et_al_2006,lidz_et_al_2008,zahn_et_al_2011,pober_et_al_2014}.
(It is worth noting that these earlier simulations were run under the assumption of
$T_S \gg T_{\rm CMB}$, but we find in our current simulations
that the neutral fraction remains the principal
factor in determining the shape of the 21~cm
power spectrum even when this assumption is relaxed.)

\subsection{Modelling Framework}

The basic approach of this work is to explore the 21~cm power spectra
that occur in the two-dimensional parameter space of $(T_s, x_{\rm HI})$.
Paper I explores this parameter space using a toy, analytic model for 
the ionization fluctuation power spectrum from ``patchy" reionization: 
$\Delta^{2}_{i}(k) = (x_{\rm{HI}} - x_{\rm{HI}}^{2})/\ln{(k_{\rm max}/k_{\rm min})}$,
and scaling its amplitude to model the effects of a globally cold spin temperature 
\citep{parsons_et_al_2012a,parsons_et_al_2014}.
We undertake a more physically motivated mapping of this space, using the publicly available
\texttt{21cmFAST}\footnote{http://homepage.sns.it/mesinger/DexM\_\_\_21cmFAST.html/} 
code v1.04 \citep{mesinger_and_furlanetto_2007,mesinger_et_al_2011}.
Since each run of this code gives a full ionization history, we need to vary
only the spin temperature as a function of neutral fraction, which we
accomplish by varying the X-ray production efficiency, $\zeta_X$.
By effectively lowering the number of X-rays produced per stellar baryon,
we reduce the rate at which the gas is heated, allowing us to produce power
spectra for a large number of low spin temperatures.  
Thus, we use the new PAPER observations to place constraints on the relative timing of the EoR and 
X-ray heating epochs: the IGM can not be too cold outside of the cosmic HII patches, 
because the resulting 21~cm power would be too large at $z\approx8.4$.   
Our methodology strives to quantify this statement, using fiducial EoR and X-ray heating models
(all parameters other than $\zeta_X$ in \texttt{21cmFAST} are kept at their default fiducial values)
and varying their relative timing.

In our simulations, we assume that the galactic X-ray luminosities follow a 
power law with a spectral energy index of 1.5, down to photon energies of $h\nu\geq 0.2$ keV. 
These values are consistent with Chandra observations of local, star-forming galaxies 
\citep{mineo_et_al_2012a,mineo_et_al_2012b}, whose soft X-ray luminosities 
(relevant for heating the IGM) have comparable contributions from hot ISM emission and 
HMXBs.  \cite{pacucci_et_al_2014} recently showed that the shape of the SED of 
early galaxies can impact the large-scale 21~cm signal during X-ray heating, 
by up to a factor of $\sim$ few.  As explained below however, 
here we are most sensitive to the relative timing of the EoR and X-ray heating epochs, 
which can impact the large-scale 21~cm power by factors of $\sim$100 
\citep{christian_and_loeb_2013,mesinger_et_al_2014}.

We use \texttt{21cmFAST} to calculate the ionization histories and associated power
spectra for twelve values of $\zeta_X$ between $0$ to $2\times10^{56}$ (roughly $0$ to $0.4$
X-ray photons per stellar baryon).  Each run produces 82 redshift outputs spanning 
the range $z = 6.14$ to $z = 34.51$.  
We then interpolate power spectra across the 3D space $(T_S, x_{\rm HI}, k)$.
However, these simulations do not regularly cover this space; since the simulations
start at such high redshift, there are significantly more data points at high neutral fractions
than low.  However, in the range where the PAPER constraints are significant, we have
data points spaced by $\sim 0.6$~K in $T_S$ and $\sim 0.05$ in neutral fraction.

\begin{figure*}
\centering
\includegraphics[width=6.5in]{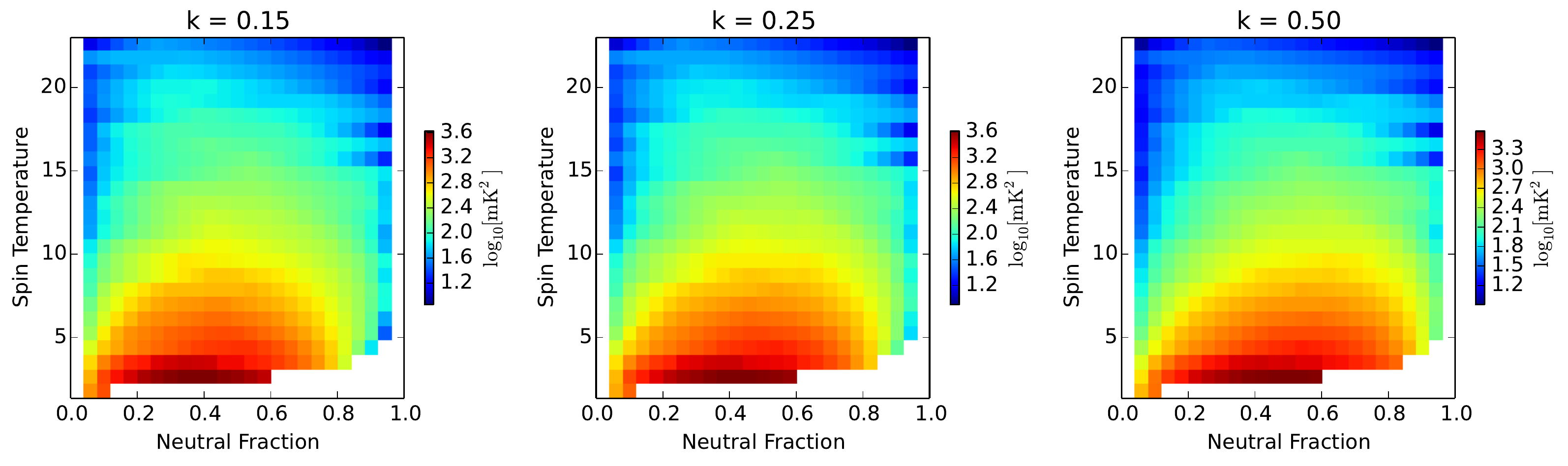}
\caption[]{Power spectrum values in log$_{10}~\mathrm{mK}^2$ for different 
combinations of $T_S$
and $x_{\rm HI}$ at $z = 8.4$; 
the three panels show the value at $k = 0.15,\
0.25,\ \mathrm{and}\ 0.50~h\rm{Mpc}^{-1}$, respectively.  It is clear
that the power spectrum is relatively flat in $k$, with the brightest values
occurring at very low spin temperatures.
The sharp drop off in power above $T_S = 20$~K
is expected; these spin temperatures are approaching the CMB temperature at $z=8.4$, 
and as can be seen
from Equation \ref{eq:deltaTb} for the brightness temperature, 
there is little to no 21~cm signal when the spin
and CMB temperatures are comparable.$^{\ref{footnotenumber}}$}
\label{fig:kslices}
\end{figure*}

Slices through this space at $k = 0.15$, 
$0.25$, and $0.50~h\rm{Mpc}^{-1}$ are shown in Figure \ref{fig:kslices}.
As expected, the brightest power spectra occur at low spin temperatures. 
The drop-off in ampltiudes at high and low neutral fractions is also straightforward to
understand.  At high neutral fractions, the spatial distribution of $x_{\rm HI}$ is 
relatively uniform, and this term does not add appreciable power to the power spectrum;
the predominant contribution comes from density ($\delta$) fluctuations, which have 
a smaller amplitude.
At low neutral fractions, the lack of neutral hydrogen leads to an overall absence of the 21~cm 
signal.  At the mid-point of reionization ($x_{\rm HI} \approx 0.5$), however, large ionized bubbles
constrast strongly with predominantly neutral regions, leading to large spatial fluctuations
in the $x_{\rm HI}$ field and a large 21~cm power spectrum amplitude.

One of the motivations for using such an approach is the elimination of
redshift as a parameter to be explored.\footnote{In 
order to correct for the effects of redshift --- \texttt{21cmFAST}
produces ionization histories where each neutral fraction corresponds to 
a specific redshift, whereas we want to compare to measurements specifically at $z = 8.4$ --- 
we scale each power spectrum relative to the CMB temperature
at $z = 8.4$:
$P(k,z = 8.4) = P_{\texttt{21cmFAST}}(k,z) * \left(\frac{T_S - T_{\rm CMB}(z = 8.4)}{T_S - T_{\rm CMB}(z)}\right)^2.$
}
While one of the principal
goals of early 21~cm experiments is to determine the evolution of the cosmic
neutral fraction as a function of redshift, the level of the current PAPER
upper limit does not allow for such an analysis.  And, while the PAPER
measurement is at only one redshift, this method allows us to use that
measurement to place constraints on the spin temperature without knowing
the neutral fraction at $z=8.4$.  
The other main benefit of this approach is that it eliminates the need
to run simulations which only vary what are effectively ``timing" parameters in the
\texttt{21cmFAST} code: parameters that change the redshift
at which specific neutral fractions occur, while having little to no
effect on the shape or amplitude of the power spectrum at fixed
neutral fraction.
\footnotetext{Some interpolation artifacts can also be seen at low and high
neutral fractions for the larger spin temperatures; these features occur where the \texttt{21cmFAST}
simulations give few data points.  These artifacts have no
effect on the conclusions of this work, which are concerned
with the brighter power spectra at lower $T_S$ where the space is well-sampled.\label{footnotenumber}} 

\subsection{Effect of Other Parameters}
\label{sec:otherparams}

The exact shape and amplitude of the 21~cm power spectrum is the result of a
rich combination of astrophysics and cosmology, which is difficult to fully
map out in any parameter space, let alone a three-dimensional one.  
In this section, we consider the effects of parameters other than the spin temperature
and neutral fraction on the 21~cm power spectrum.
When the IGM is cold relative to the CMB, we find that these other effects are sub-dominant, 
amounting to relatively small corrections to our quantitative results.
The reader primarily interested in the IGM temperature limits placed by PAPER
can skip to \S\ref{sec:results}.

In simulations where $T_S \gg T_{\rm CMB}$,  
the properties of the sources contributing to reionization 
are the dominant drivers of the 21~cm power spectrum and its evolution.
In the \texttt{21cmFAST} code,
these parameters include $\zeta$, the ionizing
efficiency of galaxies --- which we find has no effect on the shape or amplitude
of the power spectrum --- and the minimum virial temperatures of the halos
that can produce both ionizing or X-ray photons --- which has a small effect
on the power spectrum relative to the global spin temperature, but
predominantly changes the timing of reionization and/or heating
\citep{mesinger_and_furlanetto_2007,mesinger_et_al_2011,pober_et_al_2014}.
Simulations which spanned the range of reasonable values for the minimum virial
temperature of ionizing halos in \cite{pober_et_al_2014} only varied
the peak power spectrum brightness at 50\% ionization between $\approx 10-30~\rm{mK}^2$.
This scale is far below the power spectrum amplitudes achievable with a cold
IGM, which can range from several hundred to several thousand mK$^2$,
as seen in Figure \ref{fig:kslices}.

\begin{figure*}
\centering
\includegraphics[width=6.5in]{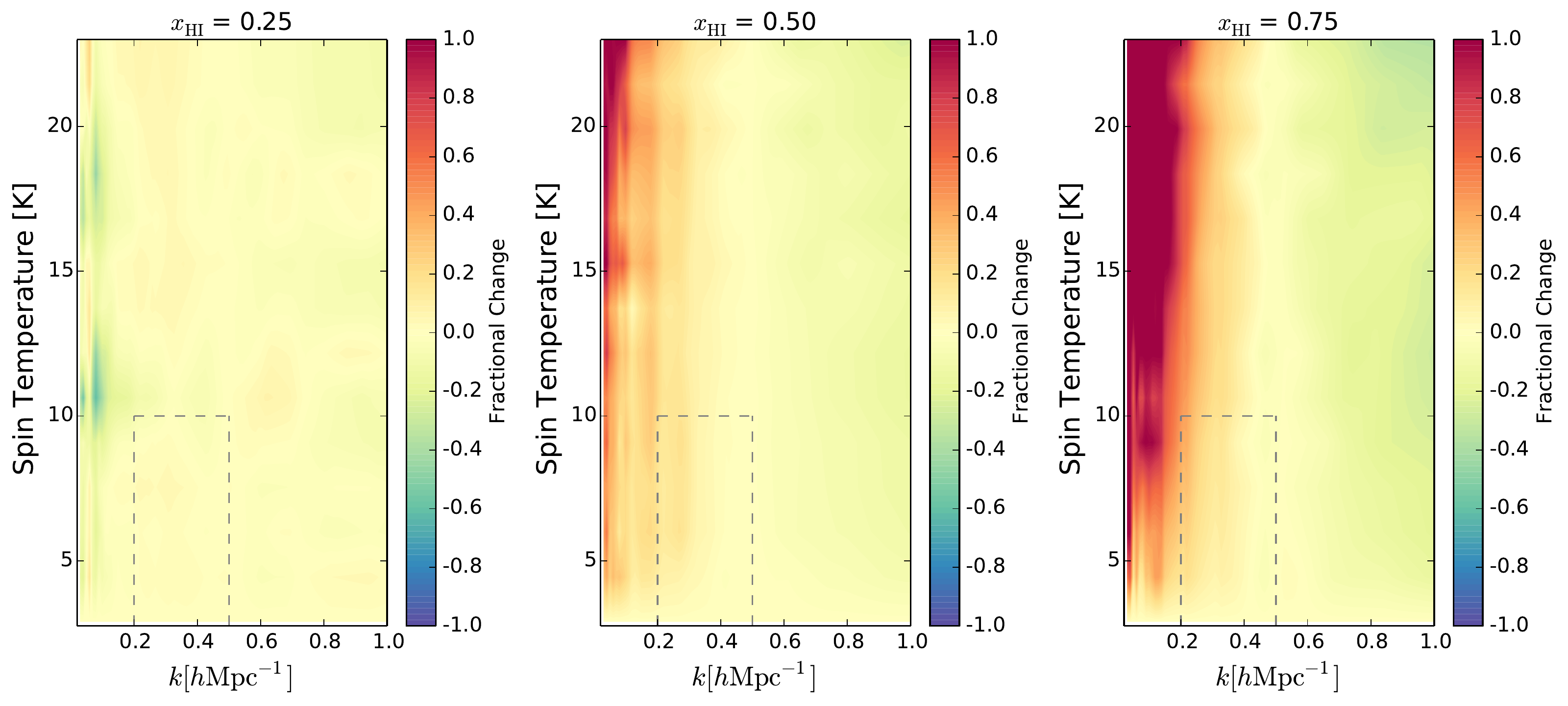}
\caption{The fractional change in the 21~cm power shape caused by spin temperature fluctuations
as a function of global spin temperature (i.e. heating efficiency).  The three panels
show different neutral fractions $x_{\rm HI} = 0.25$ (\emph{left}), 
$x_{\rm HI} = 0.5$ (\emph{center}), and $x_{\rm HI} = 0.75$ (\emph{right}).
In the regime where the PAPER constraints apply 
($0.2~h\rm{Mpc}^{-1} < k < 0.5~h\rm{Mpc}^{-1}$, $T_S < 10$~K, marked with the dotted lines),
varying spin temperature fluctuation morphology 
changes the power spectrum shape by $\sim$ ten percent or less.}
\label{fig:errorcalc}
\end{figure*}

Driving the effective dominance of the spin temperature in setting the
amplitude of these strong ($\sim$\ thousand mK$^2$) 21~cm signals 
is the contrast of the cosmic ionized patches ($\delta T_b \approx 0$ mK) with the cold 
($\delta T_b \gtrsim -200$ mK), neutral patches 
\citep{christian_and_loeb_2013,mesinger_et_al_2014,parsons_et_al_2014}.  
We note however that the uncertainty 
in the precise morphological structure of the ionization and temperature 
fluctuations does quantitatively 
impact our constraints.
Our approach largely assumes that the contribution of spatial fluctuations in the spin temperature
to the 21~cm power spectrum does not vary with the effectiveness of X-ray heating.  One expects
that spin temperature fluctuations would be small if heating is inefficient and would grow
larger with stronger heating.  We quantify the change in the shape of the 21~cm power spectrum
versus heating efficiency in Figure \ref{fig:errorcalc}.
For low neutral fractions, the change is small ($\sim 10\%$) regardless of
spin temperature.  At higher neutral fractions, spin temperature fluctuations increase
the 21~cm power spectrum at large scales ($k < 0.2~h\rm{Mpc}^{-1}$) by factors of as much as
3 when heating is relatively efficient and the global spin temperature is high.  We also see
a decrease in 21~cm power by factors of up to $\sim30\%$ at smaller scales ($k > 0.6~h\rm{Mpc}^{-1}$).
However, in the regime constrained by the PAPER measurements 
(scales $0.2~h\rm{Mpc}^{-1} < k < 0.5~h\rm{Mpc}^{-1}$, 
and, as shown subsequently, spin temperatures less than 10~K, demarcated by the dotted lines), 
we find spin temperature fluctuations
introduce typical changes of $\sim 10\%$ or less.
Therefore, our approximation of a spin temperature field morphology independent of heating
efficiency should introduce only small uncertainties into our constraints.
This result largely confirms the findings of 
\cite{pritchard_and_furlanetto_2007}, and validates the approach
of Paper I to use the global spin temperature as a multiplicative scalar for the overall
power spectrum amplitude.

Our methodology also implicitly assumes that the contribution to the overall power spectrum
from the density, velocity, and 
temperature fluctuation terms
that contribute to $\Delta^2(k)$ changes minimally over the redshift
range spanned by the simulations.  The ionization history simulated by \texttt{21cmFAST} is
in principle
independent of the heating history; in our simulations, the IGM is 90\% neutral at $z=12.5$,
50\% neutral at $z=9.5$, and 10\% neutral at $z=8$.  
In extrapolating all neutral fractions to $z=8.4$, we have assumed that the ionization
field provides the dominant contribution to the power spectrum across this redshift range,
and that the fractional contribution of the density and velocity
fields to the power spectrum evolves relatively slowly.  
This is in general a good assumption
for neutral fractions between $\approx10\%$ and 90\%, which are achieved over a narrow
redshift range, but makes the interpretation of
high and low neutral fractions (where our constraints are the poorest) more questionable.
We do expect the spin temperature fluctuations to grow more important at lower redshifts
(the temperature of an overdense region grows faster than its density), but this is also a relatively
small effect over the redshift range we extrapolate from. 

One additional free parameter in \texttt{21cmFAST} is 
the mean free path of ionizing photons through the IGM, 
which primarily accounts for the unresolved,
self-shielded pockets of neutral gas that limit the extent of HII regions.
This parameter has been shown to alter the shape of the 21~cm power spectrum
\citep{sobacchi_and_mesinger_2014,pober_et_al_2014,greig_and_mesinger_2015},
but principally only on the largest scales, at which the PAPER measurements
are limited by residual foreground emission.

These caveats do suggest that our quantitative results should not be too strictly interpreted,
as we have neglected several effects that could change the constraints by $\sim$ tens of
percent.
Given the scale of the current PAPER upper limit and the range of $k$ modes measured, 
however, working in the two-dimensional parameter space of spin temperature and neutral fraction
remains a well-motivated approach.

\hspace{0.5in}
\section{Results}
\label{sec:results}

At each position in the $(T_S,x_{\rm HI})$ space plotted in Figure \ref{fig:kslices},
we calculate the probability of getting the measurements shown in Figure \ref{fig:data_pspec}
given our model \texttt{21cmFAST} power spectrum at those values of $(T_S,x_{\rm HI})$.  
We calculate the joint likelihood across all values of $k$ measured by
PAPER.  As described in Paper I,
the $2\sigma$ error bars plotted in Figure \ref{fig:data_pspec} are calculated from bootstrapping;
here, we assume they follow a Gaussian distribution to allow for analytic calculation of the 
likelihood.
We also make the conservative choice to treat all our measurements as upper limits on the
21~cm signal so that we only exclude models which predict \emph{more} power than we observe.
The PAPER measurements clearly detect non-zero power at several wavemodes;
if treated as detections of the 21~cm signal, these points would exclude, e.g., the model
power spectrum in Figure \ref{fig:data_pspec} for being too faint.
Therefore, when calculating the likelihood that our data are consistent
with a given model, we exclude any constraints from points brighter than the model prediction.

\begin{figure}
\centering
\includegraphics[width=3.5in]{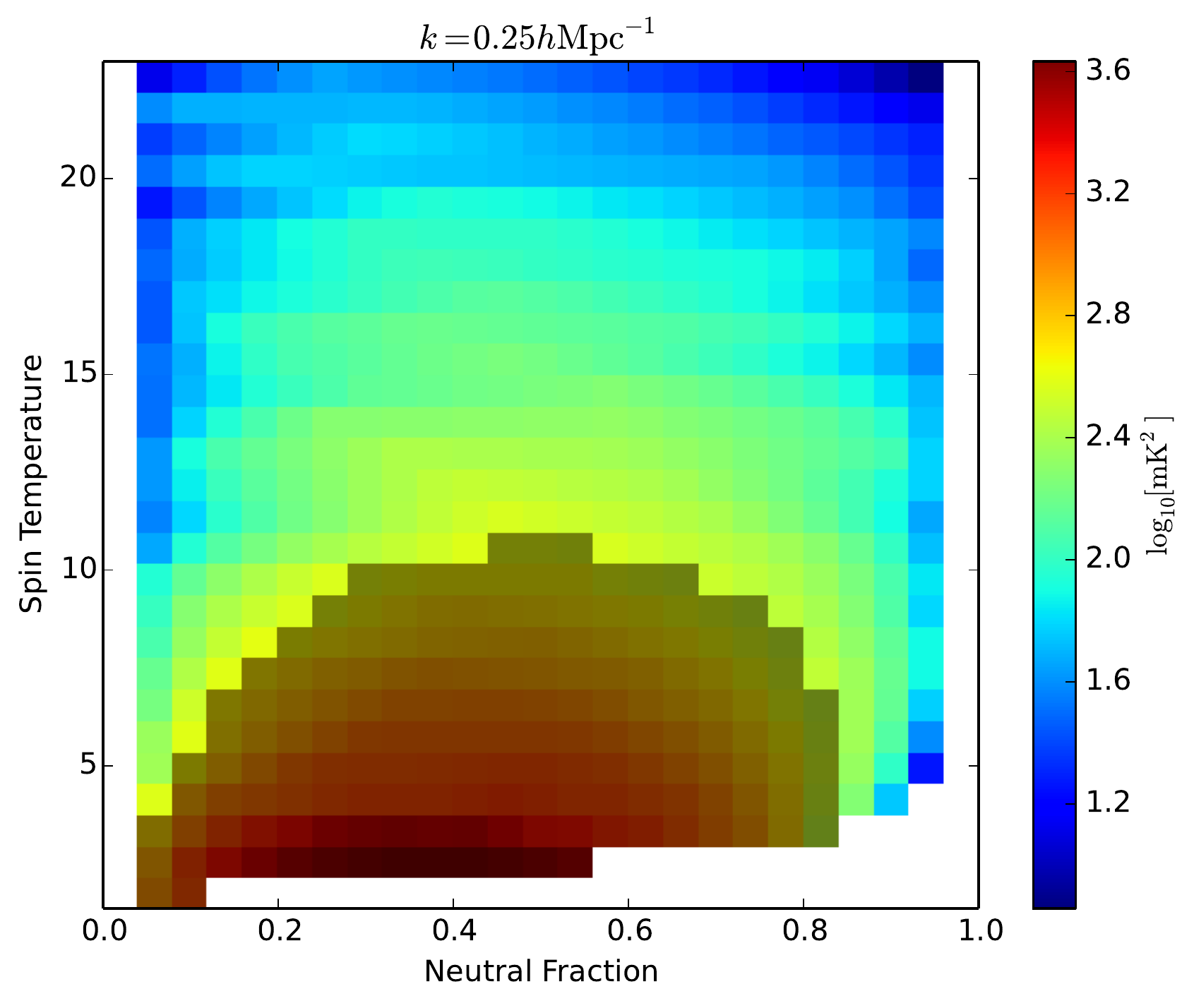}
\caption{
Constraints
on the IGM spin temperature as a function of neutral fraction
based on the 2$\sigma$ upper limits from the PAPER measurements; regions
excluded at greater than 95\% confidence are shaded in gray.
Plotted is a slice through our 3D $(T_s, x_{\rm HI}, k)$ space at the $k=0.25~h\rm{Mpc}^{-1}$,  
but the constraints are calculated from the joint likelihood across all $k$
modes measured by PAPER. 
}
\label{fig:constraints}
\end{figure}

The constraints produced by this analysis are shown in Figure \ref{fig:constraints}.
As expected, our measurements are inconsistent with very low spin temperatures, as these models
produce the brightest power spectra.  The exact spin temperatures ruled out by our data
depend somewhat on the (currently unknown) neutral fraction in the IGM at $z = 8.4$.
For neutral fractions between 10\% and 85\%, we can rule out spin temperatures below $\approx~5$~K at 95\%
confidence.  By narrowing the range of neutral fractions, our constraints grow more stringent:
if the Universe is between 15\% and 80\% neutral at $z = 8.4$, we rule out spin temperatures below
$\approx~7$~K, and for neutral fractions between 30\% and 70\%, we require $T_S$ to be greater than 
$\approx~10$~K.
The explanation for the poorer constraints at the highest and lowest neutral fractions
is straightforward: spatial fluctuations in the ionization field are small at the beginning and end
of reionization, lowering the amplitude of the power spectrum, and allowing for a colder
IGM to still be consistent with the data.
If there is no heat injection into the Universe, cosmological adiabatic cooling brings the gas 
temperature to 1.18~K at $z=8.4$ (if thermal decoupling of the gas occurs at
at $z=200$).  Assuming the Wouthuysen-Field effect has efficiently coupled
the spin temperature of the hydrogen to the kinetic temperature of the gas, our measurements
require a gas temperature $\approx 5-10$ times larger than the minimum allowed by adiabatic cooling.

Comparing these results with those from the analytic toy model calculations from Paper I,
we see general agreement, although our constraints are somewhat stronger.  
The calculation in Paper I does not include any of the physical effects of reionization;
rather, it assumes a flat power spectrum between a minimum and maximum wavenumber
and uses the integral constraint for a patchy reionization that 
$\int \mathrm{d}\log k\ \Delta^2_i(k) = x_{\rm HI} - x_{\rm HI}^2$.
The minimum and maximum wavenumbers roughly correspond to a range of ionized bubble scales;
as they grow closer together,
more power accumulates within a narrow range of wavenumbers. 
While it is difficult to directly compare this simple
model to our semi-analytic results, it is reassuring to find results constraining
a similar range of $T_S$.

\section{Discussion}
\label{sec:discussion}

It is interesting to compare our constraints on the temperature of the $z=8.4$ IGM with
theoretical predictions.  As stated previously, most models of early galaxy formation
and reionization predict fairly efficient heating, such that the temperature
of the IGM is much greater than that of the CMB by $z \lesssim 10$
\citep{furlanetto_2006,mcquinn_and_oleary_2012}.
Here, we consider two models that have the potential to result in low
amounts of cosmic heating: an observationally based model, where we consider 
the heating produced by the currently observed high redshift galaxy population, 
and a more physical model, where we compare with predictions 
using the recently proposed reionization model from \cite{robertson_et_al_2015} (hereafter, R15).
These two models are discussed in \S\ref{sec:obscomp} and \S\ref{sec:robertsoncomp}, respectively.

We can estimate the heating we would expect by solving the differential
equation demanded by energy conservation in an expanding universe:
\begin{equation}
\frac{\mathrm{d}T_K}{\mathrm{d}t} = -2H(z)T_K + \frac{2}{3}\frac{\epsilon_X}{k_Bn},
\label{eq:heat_diffeq}
\end{equation}
where $T_K$ is the kinetic temperature of the IGM, 
$k_B$ is Boltzmann's constant, $n$ is the number density of neutral gas particles,
and $\epsilon_X$ is the energy injected into the IGM per second per unit
volume by X-ray sources \citep{furlanetto_et_al_2006}.  To estimate 
$\epsilon_X$, we use the star formation rate densities $(\dot{\rho}_{\rm SFR}$) 
measured from the 
observed high redshift galaxy population by \cite{bouwens_et_al_2014}
and \cite{mcleod_et_al_2014}
and the local correlation between star formation and X-ray luminosity:
\begin{equation}
L_X = 3.4 \times 10^{40} f_X \frac{\mathrm{SFR}}{1\mathrm{M}_\odot \mathrm{yr}^{-1}} \mathrm{erg\ s}^{-1},
\label{eq:Lx}
\end{equation}
where SFR is the star formation rate, and $f_X$ is an unknown high redshift normalization factor
\citep{grimm_et_al_2003,ranalli_et_al_2003,gilfanov_et_al_2004}.
Equation \ref{eq:Lx} can be related to $\epsilon_X$
by $\epsilon_X = 3.4\times10^{40}f_Xf_{\rm{abs}}\dot{\rho}_{\rm SFR}$, 
where $f_{\rm{abs}}$ is the fraction of
total X-ray emission that deposits heat into the IGM.
Given $f_X$, $f_{\rm{abs}}$, and
$\dot{\rho}_{\rm SFR}(z)$, we can solve
Equation \ref{eq:heat_diffeq} to predict the temperature of the IGM.
\begin{figure*}[htbp!]
\centering
\includegraphics[width=7in]{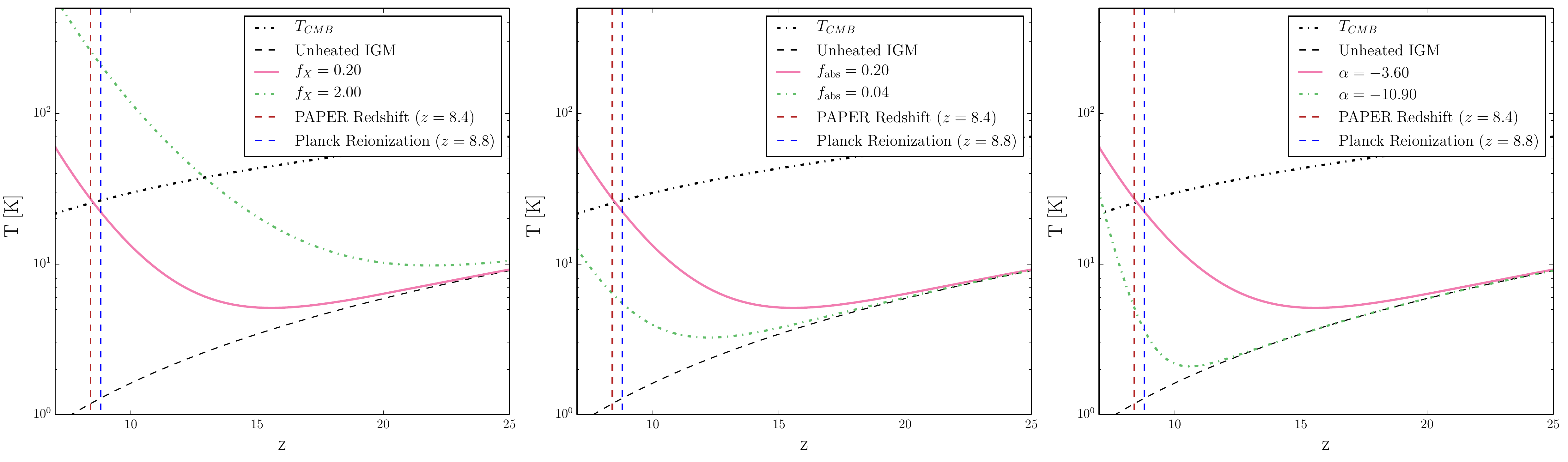}
\caption{
The dependence of evolution of the IGM temperature as a function of redshift on the uncertain 
parameters
in the calculation: $f_x$ (\emph{left}), $f_{\rm abs}$ (\emph{middle}), and $\alpha$ (\emph{right}).
In all panels, the solid pink curve plots our fiducial model with 
$f_X = 0.2,\ f_{\rm abs} = 0.2,$ and $\alpha = -3.6$.
The dashed vertical red and blue lines show the redshift of the PAPER measurement $(z=8.4)$
and the \emph{Planck} maximum likelihood redshift for an instantaneous reionization
$(z = 8.8^{+1.7}_{-1.4}$), respectively.  The black dashed curve shows the
minimum gas temperature from adiabatic cooling, and the black dot-dashed curve shows the CMB
temperature.
All predictions are based on the assumption that the observed $z = 7$ galaxy population is
providing the X-rays that heat the IGM.
}
\label{fig:heatcalc_param}
\end{figure*}

\subsection{Parameter Uncertainties}

Each of the three parameters described above is not well-determined at high redshift.  Here we discuss
the uncertainties in each before constructing models to span the range of uncertainties.

\begin{itemize}

\item \emph{$f_X$.}  
Early measurements of the local star formation rate/X-ray luminosity
correlation yielded, a value of $3.4 \times 10^{40} f_X \frac{\mathrm{SFR}}{1\mathrm{M}_\odot \mathrm{yr}^{-1}} \mathrm{erg\ s}^{-1}$ \citep{grimm_et_al_2003,gilfanov_et_al_2004}; 
$f_X$ is a correction factor relative to these initial measurements.
More recent measurements have suggested that the local $f_X$ value is
closer to 0.2 \citep{lehmer_et_al_2010,mineo_et_al_2011,mineo_et_al_2012a,mineo_et_al_2012b}, 
while the high-redshift value still has large uncertainties
\citep{dijkstra_et_al_2012}.  
\cite{fragos_et_al_2013} model the redshift evolution of the 
HMXB population, and find that $f_X$ may be up to 10 times higher than local,
i.e., $f_X = 2.0$,
while observations from \cite{cowie_et_al_2012} find little to no redshift evolution
in $f_X$ up to $z\approx6$.
In this work, we consider $f_X$ values which span the range $0.2 - 2.0$.

\item \emph{$f_{\rm abs}$.}
X-rays heat the IGM by first photoionizing a neutral atom. 
The resultant fast electron then deposits a fraction of its energy
into heating the IGM gas itself.  The relative fraction of 
this ionization energy which goes into heating 
depends on the X-ray photon energy and the neutral fraction of the
hydrogen gas \citep{shull_and_van_steenberg_1985,chen_and_kamionkowski_2004,valdes_and_ferrara_2008,furlanetto_and_stoever_2010}.
However, since we are concerned only with the temperature of the predominantly 
neutral regions in the IGM (as this gas is responsible for the 21~cm signal), the global ionization
fraction is not the relevant quantity for determing the fraction of energy that goes to heating
the gas.  Rather, it is the small ionization fraction produced by X-rays which have penetrated
into the neutral regions; in the limit of very small ionization fractions, a constant
value of $f_{\rm abs} = 0.2$ is a good approximation \citep{valdes_and_ferrara_2008}.\footnote{The
fact that we are considering cold reionization scenarios ensures the amount of ionization caused
by X-rays must be small.  For predominantly neutral gas, the ratio of X-ray energy injection
contributing to ionization is no more than $\sim3$ times that contributing to heating
\citep{valdes_and_ferrara_2008}.  An injection of $\sim10^{-2}$~eV/atom as heat would raise
the gas temperature to $\sim 100$~K, well outside of the cold IGM regime constrained by
the PAPER measurements.  Therefore, a maximum of $\sim3\times10^{-2}$~eV/atom could have
gone into ionizations, which could produce an ionization fraction no larger than $\sim2\times10^{-3}$.
}

Recently, \cite{fialkov_et_al_2014} suggested that previous heating calculations assumed
too soft a spectra for X-ray emitters (specifically, HMXBs) at high redshift.
Using a spectral model with fewer soft photons, they find the absorption of X-rays by the
IGM can be reduced by as much as a factor of 5.  
\cite{pacucci_et_al_2014} argue the opposite:
that the early galaxies have softer X-ray spectra due to: 
(i) a contribution from the thermal emission from the hot interstellar medium (ISM), 
which is locally found to be comparable at the relevant energy range to that of the HMXBs 
(only the latter is used to motivate the scaling in Equation \ref{eq:Lx}); 
and (ii) lower column densities and metallicities of early galaxies compared to local ones, 
resulting in more soft X-rays escaping into the IGM.  
For robustness, we explore a broad parameter space, considering values of $f_{\rm abs}$
spanning the range $0.04 - 0.2$ in our calculations.

\item \emph{Star Formation Rate Density.}
There has been some debate in the literature as to the evolution of the star formation
rate density at high redshift.  Modeling the star formation rate density redshift evolution
as a power law, $\dot{\rho}_{\rm SFR} \propto (1+z)^{\alpha}$, \cite{bouwens_et_al_2014} and
\cite{oesch_et_al_2014} find a very steep drop off of $\alpha=-10.9$ at high redshift.
Other measurements from \cite{mcleod_et_al_2014}
find a much shallower evolution, $\dot{\rho}_{\rm SFR} \propto (1+z)^{-3.6}$ (although the best fits
come from a more complex functional form with a slightly steeper slope).
We consider models using both these power law indices, although for our fiducial model
we choose $\alpha = -3.6$, since a larger high redshift 
$\dot{\rho}_{\rm SFR}$ seems
necessary to produce the \emph{Planck} optical depth (\citealt{planck_2015_13}, R15).
To set the overall scale of this power law, we set the star formation rate density at $z=7$ to be 
$10^{-2}~\mathrm{M_\odot Mpc^{-3} yr^{-1}}$, consistent with the observed
values from \cite{bouwens_et_al_2014} and \cite{mcleod_et_al_2014}
(although a correction for extinction does result in slightly higher values).

\end{itemize}

The effect on the heating history from varying each parameter 
is plotted in Figure \ref{fig:heatcalc_param}; the left
hand panel shows the effect of changing $f_X$, the middle $f_{\rm abs}$, and the right $\alpha$.
The parameters not being varied are held at fiducial values of $f_X = 0.2,\ f_{\rm abs} = 0.2,$ and
$\alpha = -3.6$; in each panel, this fiducial model is plotted in solid pink.
Also plotted is the expected gas temperature in the absence of any heating (black, dashed)
and the CMB temperature (black, dot-dashed).  Dashed vertical lines show the redshifts
of the PAPER measurement ($z=8.4$; red) and the \emph{Planck} maximum likelihood
instantaneous reionization redshift 
($z = 8.8^{+1.7}_{-1.4}$; blue).
The CMB temperature at $z=8.4$ is $25.4$~K; if the gas temperature is below the CMB temperature, 
then the 21~cm signal will be seen in absorption.  We see that relative to our fiducial model,
only an increased value of $f_X$ heats the gas well above the CMB temperature,
while choosing the minimum values for either $f_{\rm abs}$ or $\alpha$ can result in an IGM
temperature well below the $\approx~10$~K lower limit at $z = 8.4$ from the PAPER measurements.
\begin{figure*}[htbp!]
\centering
\includegraphics[width=3.25in]{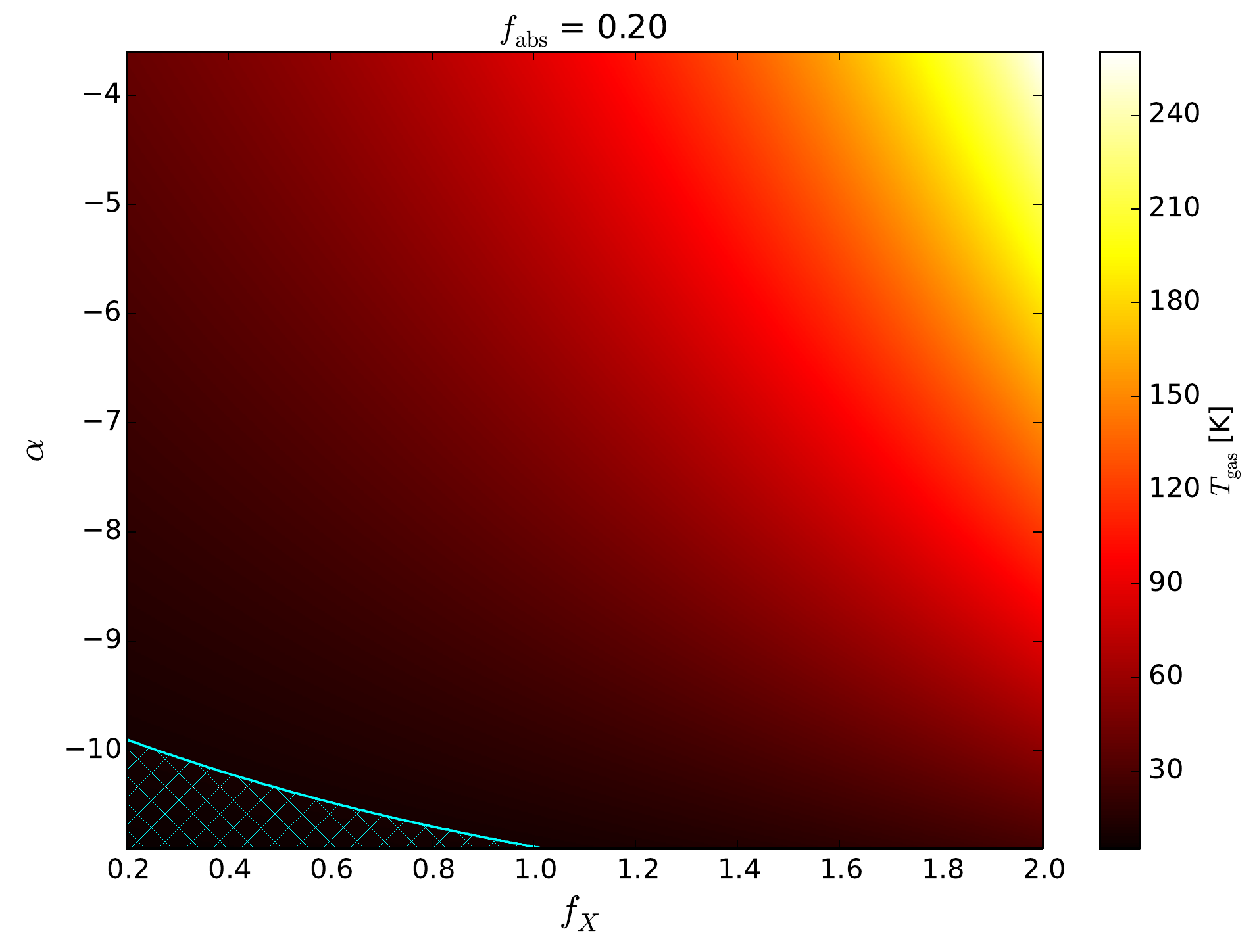}\includegraphics[width=3.25in]{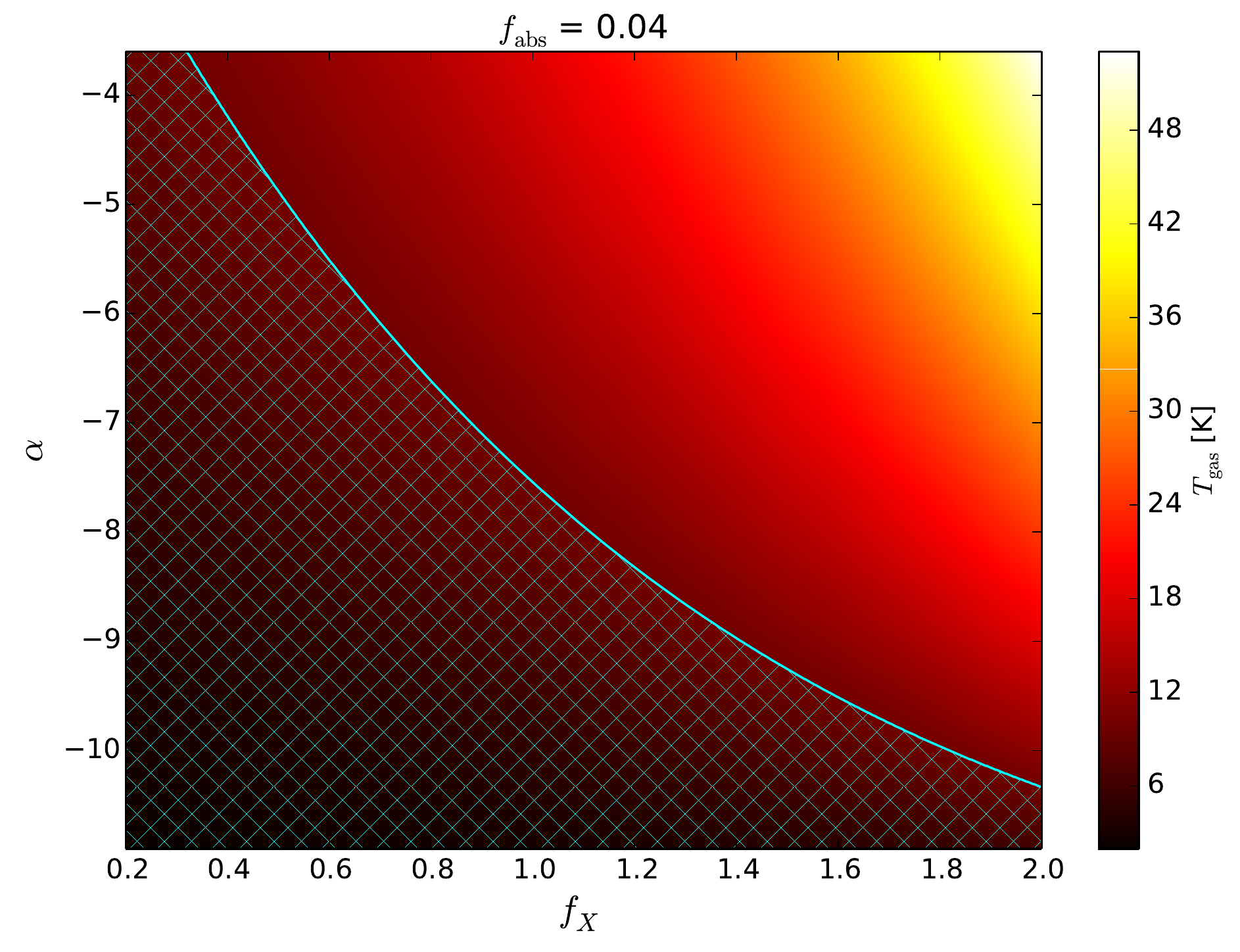}
\caption{Predicted IGM temperature produced by the observed galaxy population as a function of
the model parameters $\alpha$ and $f_X$.  The hashed cyan regions are excluded based
on the PAPER constraints that $T_{\rm IGM}~\gtrsim~10$~K (assuming the neutral fraction is between
30\% and 70\% at $z=8.4$).
\emph{Left}: $f_{\rm abs} = 0.20$.  With the fiducial absorption coefficient, very little
parameter space is ruled out.  However, we disfavor a combination of a steep slope
to the $\dot{\rho}_{\rm SFR}(z)$ relation and lower value of $f_X$ close to the locally
measured value.
\emph{Right}: $f_{\rm abs} = 0.04$, as suggested by \cite{fialkov_et_al_2014}. 
A considerable amount
of parameter space in the observed galaxy model is ruled out, disfavoring the local value of
$f_X = 0.2$ and the steeper slope of $\dot{\rho}_{\rm SFR}(z)$ of $\alpha=-10.9$.
Even with this very low value of $f_{\rm abs}$, there is still room for the observed galaxies
alone to heat the IGM well above $T_{\rm CMB}$.
}
\label{fig:constraints2d}
\end{figure*}

\hspace{0.5in}
\subsection{Comparison with Observations}
\label{sec:obscomp}

To better explore the parameter space, we plot the predicted $z=8.4$ IGM temperature
produced by the observed galaxy population versus both $\alpha$ and $f_X$ in Figure
\ref{fig:constraints2d}.
For the fiducial value of $f_{\rm abs} = 0.2$ (left), we rule out very little parameter space,
but do disfavor the combination of a steep $\alpha$ and low $f_X$.  In general, however,
the observed galaxies heat the IGM beyond both the realm ruled out by the PAPER measurements
and the CMB temperature of 25.4~K.  With a lower value of $f_{\rm abs} = 0.04$, chosen to represent
the harder X-ray spectra suggested by \cite{fialkov_et_al_2014}, we rule out a significant
region of parameter space (right panel of Figure \ref{fig:constraints2d}.)
In this model, both the steepest slopes for models
of $\dot{\rho}_{\rm SFR}$ and the smallest values of $f_X$ are excluded.
However, there is still a large range of parameter space where the IGM temperature
is heated above the limits set by PAPER.

Of course, the constraints we have placed on the parameters in our heating model
can still be avoided by invoking other sources of heating --- whether from
an additional population of fainter, yet-undetected galaxies, or from other mechanisms
such as active galactic nuclei and/or quasars \citep{volonteri_and_gnedin_2009},
shock heating \citep{gnedin_and_shaver_2004,mcquinn_and_oleary_2012} 
or even dark matter annihilation \citep{evoli_et_al_2014}.
It is often claimed that fainter galaxies 
than those observed are necessary to reionize the IGM 
(\citealt{choudhury_et_al_2008,kuhlen_and_faucher_giguere_2012,finkelstein_et_al_2012,robertson_et_al_2013}, R15); 
our measurements also imply that fainter galaxies would be required to heat the IGM
if X-ray heating turns out to be on the inefficient side of the currently allowed
parameter space.
The fact, however, that an additional galaxy population is required to complete reionization
within the observationally allowed redshift range makes the constraints arising from 
the observed galaxy heating model used in this section 
relatively unphysical.  If there are no fainter galaxies than those observed, 
it is implausible for the Universe to be significantly ionized at $z=8.4$
(although high values of the ultraviolet photon escape fraction from galaxies at high redshift
could allow for reionization to be completed by the observed galaxy population).  
And, if the IGM is significantly neutral, the PAPER lower limit on $T_S$ becomes much weaker
(c.f. Figure \ref{fig:constraints}).
However, if we remain agnostic about the source of ionizing photons, the PAPER measurements
do place constraints on the sources heating the IGM, ruling out a range of models
of X-ray heating.

\subsection{Extrapolating the Luminosity Function}
\label{sec:robertsoncomp}

As noted above, galaxies fainter than those currently detected at high redshift are expected in
most theoretical models, and are in fact necessary to complete the reionization of the Universe
by $z = 6$ (barring other significant contributors of ionizing photons; 
\citealt{choudhury_et_al_2008,kuhlen_and_faucher_giguere_2012,finkelstein_et_al_2012,robertson_et_al_2013}, R15).
While the analysis in \S\ref{sec:obscomp}
placed constraints on the heating that can be provided by only the
observed galaxies, it is useful to explore a more self-consistent model, where we include
the heating that would be generated by the galaxies necessary to reionize the Universe.
As a model, we use the maximum-likelihood star formation fit from R15,
which uses the four-parameter fitting function of \cite{madau_and_dickinson_2014}:
\begin{equation}
\dot{\rho}_{\rm SFR}(z) = a_p \frac{(1+z)^{b_p}}{1+[(1+z)/{c_p}]^{d_p}},
\end{equation}
with best fit values of $a_p = 0.01376 \pm 0.001\ \mathrm{M_{\odot} yr^{-1} Mpc^{-3}}$, 
$b_p = 3.26 \pm 0.21$, $c_p = 2.59 \pm 0.14$, and $d_p = 5.68 \pm 0.19$.\footnote{Note, however,
that there exist correlations between these errors, and as such, the range of star formation
rate histories allowed by R15 is smaller than one might na{\"i}vely calculate.} 
These values fit the compilation of star formation rate densities in 
\cite{madau_and_dickinson_2014}, but where each measured $\dot{\rho}_{\rm SFR}$ 
value has been increased 
to include an extrapolation to fainter galaxies.
Specifically, R15 take the measured galaxy luminosity function at each redshift
and integrate down to $0.001L_*$ to produce a total estimate of star formation rate
density at that redshift (where $L_*$ is the characteristic luminosity of a galaxy).
The value of $\dot{\rho}_{\rm SFR}$ at $z=7$ predicted by the R15 model
is not significantly higher than the currently observed rates: $0.020^{+0.0029}_{-0.0025}\
\rm{M_{\odot} yr^{-1} Mpc^{-3}}$ (compared with the value of $0.01\ \rm{M_{\odot} yr^{-1} Mpc^{-3}}$
currently observed near this redshift),
suggesting that half the galaxies necessary for reionization have already been observed.
With fainter galaxies included in calculations of the IGM ionization history,
the R15 model completes reionization by 
$z\approx6$, and produces an optical depth to reionization consistent with the \emph{Planck} value
\citep{planck_2015_13}.  The reionization history predicted by this model reaches
50\% ionization at $z\approx7.5$; at $z=8.4$, the redshift of the PAPER measurements,
this model predicts the Universe is $\approx~70\%$ neutral.  See Figure 3 of R15
for the full predicted ionization history.

\begin{figure}
\centering
\includegraphics[width=3.25in]{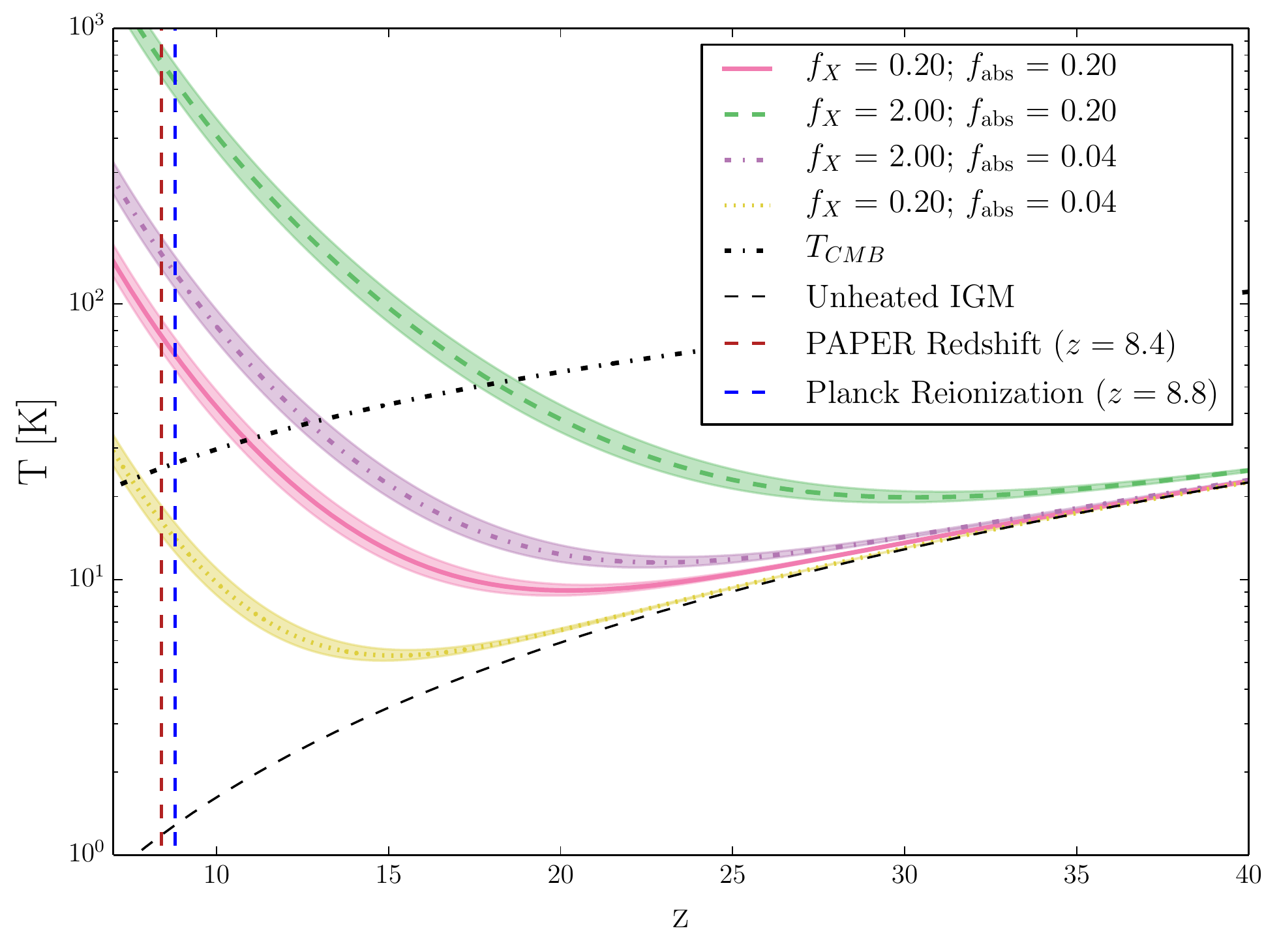}
\caption{Predicted heating of the IGM from the R15 best fit star
formation rate density evolution; our fiducial model is plotted in solid pink.
Shaded regions around each curve correspond to 1$\sigma$ uncertainties calculated from the
68\% confidence regions in the R15 model.
Even when the parameters are tuned for the weakest heating of the IGM, the model
predicts gas temperatures above those ruled out by the PAPER lower limit.
}
\label{fig:heatcalc_robertson}
\end{figure}

Figure \ref{fig:heatcalc_robertson} shows the IGM heating histories calculated from the
R15 star formation rate density evolution model for all combinations of $f_X$ and $f_{\rm abs}$
($\alpha$ is not a free parameter of the model.)
We see that even when tuned for the weakest heating ($f_X = 0.2,\ f_{\rm abs} = 0.04$),
the R15 model produces enough X-ray photons to bring the gas temperature above the
lower limit of 10~K set by PAPER.  (As argued in \cite{fialkov_et_al_2014}, this model does still
produce a relatively cold reionization, with $T_{\rm gas}~\approx~T_{\rm CMB}$ at $z\approx8$.)
Our fiducial heating model of $f_X = 0.2,\ f_{\rm abs} = 0.2$ heats the gas to $\approx~80$~K
at $z=8.4$.
While this is perhaps somewhat cooler than the $T_S \gg T_{\rm CMB}$ regime,
this contrast still amounts to only a $\approx~30\%$ decrease in the 21~cm brightness temperature,
and will be even less of an effect at $z\approx7$ where the R15 model predicts 50\% ionization.

\section{Conclusions}
\label{sec:conclusion}

We have interpreted new 21~cm power spectrum
measurements from PAPER with a semi-analytic modeling framework.  Using \texttt{21cmFAST} to cover the
parameter space of cold reionization scenarios (i.e. where the hydrogen spin temperature
is significantly below the CMB temperature during reionization), we find that power spectra with
amplitudes above $\sim 100-1000~\rm{mK}^2$ are generically produced for a wide range of neutral fractions,
effectively independent of the other physical parameters during reionization.  We cover the
2D $(T_S, x_{\rm HI})$ space with a suite of simulations, and find that the PAPER measurements rule
out spin temperatures below $\approx~5$~K for neutral fractions between 10\% and 85\%.
More stringently,
if the Universe is between 15\% and 80\% neutral at $z = 8.4$, we rule out spin temperatures below
$\approx~7$~K, and for neutral fractions between 30\% and 70\%, we require $T_S$ to be greater than 
$\approx~10$~K.
Given the recent measurements from \emph{Planck}, which suggest the midpoint of reionization occurs
at $z = 8.8^{+1.7}_{-1.4}$ \citep{planck_2015_13}, it is probable that the stronger
lower bound of $T_S > 10$~K applies.  (Using an extended model of reionization,
R15 predict the Universe to be $\approx 70\%$ neutral at $z=8.4$, which would
slightly lower our bounds on $T_S$.)

We also explore a range of models for predicting the amount of heating the observed high-redshift
galaxy population provides by $z=8.4$.  We find that the observed galaxy population
can generally heat the gas above 10~K and, thus, is not constrained by the PAPER measurements.
However if the star formation rate density $\dot{\rho}_{\rm SFR}$
drops off very steeply with redshift
(as suggested by e.g. \citealt{bouwens_et_al_2014} and \citealt{oesch_et_al_2014})
and the correlation between X-ray luminosity and star formation rate is close to the locally
observed value of $f_X = 0.2$, we find that additional heating of the IGM beyond
that provided by the observed galaxies is required.  If the X-ray emission from high-redshift
galaxies has fewer low-energy photons than expected and the IGM is heated
less-efficiently, our constraints tighten considerably, ruling out all scenarios with either a steep
redshift evolution in $\dot{\rho}_{\rm SFR}$ or a low value of $f_X$.

Lastly, we considered the predicted X-ray heating of the IGM that would be expected
under the reionization model of \cite{robertson_et_al_2015}.  This model
produces an optical depth to the CMB in good agreement with the new \emph{Planck} measurements
\citep{planck_2015_13}, and completes reionization by $z\approx6$.  We find that for all combinations
of heating parameters, this model heats the IGM above the lower limit set by PAPER and, for our
fiducial heating model, results in a spin temperature well above the CMB temperature.
Therefore, the current lower limit on $T_S$ set by PAPER is consistent with the R15 model, but does
not 
require more high redshift star formation than that suggested by galaxy and CMB observations alone.

The potential for future 21~cm studies is also clear from Figure \ref{fig:constraints}.  With each
increase in sensitivity, each new 21~cm measurement 
will rule out more of the $(T_S,x_{\rm HI})$ 
parameter space.
Figure \ref{fig:heatcalc_robertson} suggests that a spin temperature lower
limit of $\approx~15$~K would begin to rule out low efficiency X-ray heating models
otherwise consistent with 
the R15 star formation rate density history; comparison with Figure \ref{fig:constraints}
shows that placing a lower limit of this scale should require our a power spectrum
spectrum upper limit to improve by only a factor of $\approx~2$ (in units of mK$^2$).
And, with an order of magnitude more sensitivity (again, in power units --- only a factor of $\approx~3$ in
brightness sensitivity), 21~cm studies will begin to constrain the properties of the first galaxies
\citep{pober_et_al_2014,greig_and_mesinger_2015}.  PAPER has collected data with its full 128-element
array for two seasons (double both the amount of data and number of antennas used to produce the
constraints in this work), and the Hydrogen Epoch of Reionization Array 
(HERA; \citealt{pober_et_al_2014}) has broken ground and will begin operations in a few years.  
With the continuing increase in the sensitivity of 21~cm measurements,
we can expect to learn much more about the high-redshift universe in the near future.

\acknowledgements{PAPER is supported by grants from the National Science Foundation 
(NSF; awards 0804508, 1129258,
and 1125558). JCP, ARP, and DCJ would like to acknowledge NSF support (awards 1302774, 1352519, and
1401708, respectively).  JEA would like to acknowledge a
generous grant from the Mount Cuba Astronomical Association for computing resources. We graciously thank
SKA-SA for onsite infrastructure and observing support.
In addition we would like to thank our South African interns Monde Manzini and Ruvano Casper from Durban
University of Technology (DUT), who helped build out
the array from 32 antennas to the 64 antennas this analysis was based on.  The authors would also like to thank Anson D'Aloisio, Adam Lidz, Jordan Mirocha, and
Jonathan Pritchard for very helpful and fruitful conversations, Brant Robertson for providing the quantitative error contours in the R15 model, and our reviewer for helpful suggestions for better quantifying the effects of certain approximations in our work.
}

\bibliographystyle{apj}
\bibliography{ms_arxiv}{}

\end{document}